\documentclass[iop]{emulateapj}
\usepackage[usenames,dvipsnames,svgnames]{xcolor}
\usepackage[hyperfootnotes=false, linktocpage=true, colorlinks,linkcolor=Red, citecolor=blue]{hyperref}

\usepackage{float}
\usepackage{amsmath,amstext,amssymb}
\usepackage{hyperref}
\usepackage{graphicx}

\shorttitle{Double-peaked profiles in Seyfert 1}
\shortauthors{Storchi-Bergmann et al.}
\begin{document}

\title{Double-peaked profiles: ubiquitous signatures of disks in the Broad Emission Lines of Active Galactic Nuclei}

\author{T.\ Storchi-Bergmann\altaffilmark{1,3}, J.S.\ Schimoia\altaffilmark{1,2},  B.M.\ Peterson\altaffilmark{2,3}, M. Elvis\altaffilmark{4}, K.D.\ Denney\altaffilmark{2} , M. Eracleous\altaffilmark{5}, R. S. Nemmen\altaffilmark{6}}
\affil{$^1$Instituto de F\'isica, Universidade Federal do Rio Grande do Sul, Campus do Vale, Porto
Alegre, RS, Brazil}
\affil{$^2$Department of Astronomy, 140 West 18th Avenue, and the Center for Cosmology and
AstroParticle Physics, 191 West Woodruff Avenue, The Ohio State University, Columbus, OH 43210, USA}
\affil{$^3$Space Telescope Science Institute, 3700 San Martin Drive, Baltimore, MD 21218}
\affil{$^4$Harvard-Smithsonian Center for Astrophysics, 60, Garden Street, Cambridge, MA 02138, USA}
\affil{$^5$ Department of Astronomy \& Astrophysics, The Pennsylvania State University, University Park,
PA 16802, USA}
\affil{$^6$ Instituto de Astronomia, Geof\'{i}sica e Ci\^encias Atmosf\'ericas, Universidade de S\~ao Paulo, 05508-090 S\~ao Paulo, SP, Brazil}
\email{thaisa@ufrgs.br}

\begin{abstract}
Broad ($\sim$10,000\,km\,s$^{-1}$), double-peaked emission-line profiles of Balmer lines emitted by
active galactic nuclei (AGN) are thought to originate in the outer parts of
an accretion disk surrounding a nuclear supermassive black hole (SMBH), at $\sim$\,1000 gravitational
radii and are most frequently observed in the nuclear spectra of low-luminosity AGN (LLAGN) and radio-galaxies. 
In the present paper we argue that broad double-peaked profiles are present 
also in the spectra of other Type 1 AGN, such as Seyfert 1 galaxies, suggesting that the inner part of the broad-line region (BLR) is also the outer part of the accretion disk. 
We use the Palomar spectral survey of nearby galaxies to show that the only difference between Seyfert 1 BLR line profiles and  ``bona fide" double peakers is that, in most cases, besides a disk component, we need an additional Gaussian component attributed to non-disk clouds. 
The recognition that the inner and most variable part of the BLR has a disk geometry suggests that the
factor $f$ in the expression to obtain the SMBH mass in Type 1 AGN $M_{\rm BH} = f \left( R_{\rm BLR} \Delta V^2 / G \right)$
is $f=1/\sin^2 i$ for the disk dominated sources. Our median $i=27^\circ$ implies $f=4.5$, very close to the most recent value of $f= 4.3\pm 1.05$,
obtained from independent studies.
We derive a relation between $f$ and the FWHM of the broad profile that may help to reduce the uncertainties in the SMBH mass determinations
of AGN.
\end{abstract}


\section{Introduction}

Double-peaked emission lines in the spectra of active galactic nuclei (AGN), usually observed in the
permited H and He lines, are believed to originate in the outer parts of a disk,
typically at about 1000 gravitational radii ($R_g = GM_{\rm BH}/c^2$)
from the supermassive black hole (SMBH) of mass $M_{\rm BH}$. This line-emitting region 
is often taken to be the outer regions of the accretion disk that extends down to a few $R_g$.
A disk origin for double-peaked emission lines
was first proposed by \citet{Ch89} and \citet{CeH89} for the archetypical ``double-peaker" Arp\,102B. 
Subsequent work by \citet{EH94, Eracleous03} found a
number of other sources, confirming that a
disk (or ring) model does reproduce the double-peaked profiles. A similar conclusion was reached by
\citet{Strateva03} for a sample of 116 AGN with double-peaked Balmer
lines profiles selected from the Sloan Digital Sky Survey.

There have been other suggestions for the origin of double-peaked emission-line profiles.
\citet{Gaskell83} and \citet{Peterson87} suggested that they might be the
signature of binary black holes. \citet{Zheng90} pointed out that a doube-peaked profile could originate
in a biconical outflow, while \citet{Wanders95} and \citet{Goad96} showed
that it could be produced by an anisotropic continuum source,
or, for that matter, a highly anisotropic distribution of emission-line gas. The preponderance
of evidence, however, still points to a moderately to highly inclined disk-like structure \citep{Eracleous03}.

Some double-peaked profiles have been monitored spectroscopically over timescales of years, at typical intervals of
months. Examples are the cases of 3C\,390.3 \citep{VZ91,Sergeev02, Jovanovic10, Shapovalova10,
Popovic11, Sergeev11}, which has also been monitored on reverberation timescales
\citep{Dietrich98,Dietrich12},
NGC\,1097 \citep{SB03}, Arp\,102B \citep{Sergeev00,Shapovalova13}, as well as
14 other double-peaked emitters \citep{Gezari07,Lewis10}.
These observations have revealed that the double-peaked Balmer line profiles in these sources display
variability on timescales ranging from months
to years. The changes in the profiles are typically varying asymmetries in the blue and/or red peaks of
the profiles on the {\it dynamical timescale}, associated with the rotation of the gas or patterns in the gas surrounding the SMBH. 
In the particular case of NGC\,1097, during the ~30 years it has been spectroscopically
monitored \citep{SB03,Schimoia12}, the relative intensity of the blue and red peaks 
has alternated between a stronger blue peak and a stronger red peak on timescales of months. This beheavior has
been attributed to a spiral arm rotating in the disk, with a rotation period of about
$\sim$\,18\,months.


Most of the monitoring campaigns of {\it double-peakers} reported to date were carried out using very sparse cadences, typically one observation per month or even sparser. But recent monitoring of the nucleus of NGC\,1097 by \citet{Schimoia12} included a few observations at weekly intervals and
showed that there are variations in the total flux of the line and in the
separation between the blue and red peaks on a timescale of a week, i.e., on the reverberation timescale. In order to search for possible shorter timescale variations, and to try to identify the driving source of the
line variations, we subsequently monitored the double-peaked H$\alpha$ profile in the nuclear spectrum
of NGC\,1097 with the SOAR telescope and the {\it Swift} satelite \citep{Schimoia14}.
These observations confirmed that the integrated flux of the line and the velocity separation betwween
the blue and red peaks vary on a timescale as short as 5 days.
This timescale is compatible with the light-crossing time between the center of the disk and a typical
radius of $1000\,R_{g}$, where the emission line originates, according to our models.

The variability timescale of $\sim5$\,days, observed in NGC\,1097 is of
the order of the delays found between the variations of the continuum and broad
emission lines in reverberation mapping studies of nearby Seyfert 1 galaxies of comparable luminosity.
This indicates that the line-emitting disk 
in NGC\,1097 is at a distance from the ionizing source that is typical of the broad-line region (BLR)
of nearby Seyfert \,1 galaxies \citep{Peterson04}. 

It is also interesting that many Seyfert 1 galaxies show, at least on some occasions, evidence for an underlying disk-like component. Double peaks or “shoulders” are frequently seen in the Balmer-line profiles of highly variable and closely monitored sources like NGC\,5548 \citep{Sergeev07}. These various observations lead one to consider the possibility that the Balmer-line profiles in all Seyfert 1 galaxies have a disk-like component that could be identified with the outer parts of an accretion disk. A simple hypothesis might be that Balmer-line profiles in Seyfert 1 galaxies are a composite of: 
(1) a disk-like component, as seen clearly in the spectra of double-peaked emitters, 
and (2) another component originating in clouds that are not in the disk, e.g. in a wind component or in orbits farther out.

A model of a disk component plus another narrower broad component  to reproduce the broad Balmer line 
profiles of Seyfert 1 nuclei and quasars has been previously proposed by \citet{Popovic04} and 
collaborators \citep{Bon09,Lamura09}. A recent theoretical study by \citet{Elitzur14} also argues that the 
broad-line region of Seyfert 1 galaxies has two parts: a disk and a
system of clouds that are part of an outflowing disk wind. As the Eddington ratio drops, the high column density gas, which resides largely in the disk, dominates the emission and leads to double-peaked line profiles. A disk-wind component would be expected to be stronger at higher Eddington ratios, so this might naturally account for the relative prominence of the disk-like component in lower luminosity AGNs (LLAGNs) and the relative difficulty of identifying the disk component in more luminous, higher Eddington ratio objects.

In this contribution we study the spectra of all Seyfert\,1 galaxies of the Palomar spectroscopic survey \citep{FS85,ho95} -- selected for being an homogeneus and representative sample of galaxy nuclear spectra in the near Universe -- to show that we can identify disk-like structures in the Balmer-line profiles of many of them. An underlying disk-like component can account for much of the structure and asymmetries seen in Seyfert\,1 line profiles, and, combined with
another narrower component, attributable to other emitting clouds, can reproduce 
most of the broad-line profiles. Thus, our primary goal is to demonstrate the plausibility of the idea that the broad Balmer
line profiles of this sample Seyfert 1 galaxies can be described well by a combination of a disk-like component and a somewhat narrower, bell-shaped component.
 As the Palomar sample is a carefully selected local sample, designed to find the lowest luminosity AGNs, we also use our results to obtain statistics of the presence of the disk component, 
and discuss the implications for the calculation of the mass of the SMBHs in Type 1 AGN. This is our secondary goal.


This paper is organized as follows. In Sec. 2, we describe the disk model that we have used to reproduce 
the double-peaked profiles of the low-luminosity AGN (LLAGN) in NGC\,1097 and other
sources. In Secs. 3 and 4 we use this model to reproduce the double-peaked H$\alpha$ profiles of LLAGN and Seyfert\,1
galaxies of the Palomar survey and show the need for an additional broad Gaussian component in most cases. 
In Sec. 5, we show how the profile shape changes as a function of inclination relative to the plane of the sky. 
We use these results and the frequency of occurrence of the double-peaked profiles in  the 
Palomar sample to draw inferences about the presence of disk emission in type 1 AGN. 
In Sec. 6 we discuss the implications of our findings for the determination of SMBH masses 
in Type 1 AGN, and in Sec. 7 we present our conclusions.


\section{The  disk model}
\label{disk_model}

We use the model described by \cite{Schimoia12}, \cite{Gilbert99}, 
and \cite{SB03}, assuming that the broad double-peaked emission line originates in the outer parts of a
relativistic Keplerian disk of gas surrounding the SMBH. The
line-emitting portion of the disk is circular and located between an
inner radius $\xi_{1}$ and and an outer radius  of $\xi_{2}$, where $\xi$ is
the disk radius in units of the gravitational radius $R_g=GM/c^2$. The plane of the disk has an
inclination $i$ relative to the plane of the sky (i.e., zero degrees corresponds to a disk observed
face-on). Superimposed on the axisymmetric emissivity of the circular
disk, there is a perturbation in the form of a spiral arm. 
The corresponding  emissivity law is given by
\begin{multline} \label{arm} \epsilon(\xi,\phi) = \epsilon (\xi) \left
\{ 1 + \frac{A}{2} \exp \left [ -\frac{4 \ln2}{\delta^{2}} (\phi -
\psi_{0})^{2} \right ] \right . \\ \left . + \frac{A}{2} \exp \left [
-\frac{4 \ln2}{\delta^{2}} (2\pi - \phi + \psi_{0})^{2} \right ]
\right \},
\end{multline} where
\begin{equation}
 \epsilon(\xi) =\left\{ \begin{array}{ll} \epsilon_{0}\xi^{-q_{1}} &
 ,\,\xi_{1} < \xi < \xi_{q}\\
 \epsilon_{0}{\xi_{q}}^{-(q_{1}-q_{2})}\xi^{-q_{2}} &,\,\xi_{q} < \xi
 < \xi_{2} \end{array} \right.
\end{equation} is the axisymmetric emissivity of the disk. The
parameter $\xi_{q}$ is the radius of maximum emissivity, or saturation
radius, at which the emissivity reaches a maximum and changes behavior: $q_{1}$ is the index of
the emissivity law for $\xi_{1} < \xi < \xi_{q}$ and $q_{2}$ is the index
for $\xi_{q} < \xi < \xi_{2}$.
$A$ is the brightness contrast between the spiral arm and the
underlying disk, and the expression between square brackets represents
the decay of the emissivity of the arm as a function of the azimuthal
distance $\phi - \psi_{0}$ from the ridge line to both sides of the
arm, assumed to be a Gaussian function with ${\rm FWHM} = \delta$
(azimuthal width).

The relation between the azimuthal angle $\phi_{0}$ and the angular
position $\psi_{0}$ of the ridge of emissivity of the spiral arm is
given by
\begin{equation}
 \psi_{0} = \phi_{0} + \frac{\ln(\xi/\xi_{sp})}{\tan p},
\end{equation} where $\phi_0$ is the azimuthal angle of the spiral
pattern, $p$ is the pitch angle, and $\xi_{sp}$ is the innermost radius
of the spiral arm.

The specific intensity from each location in the disk, in the frame of
the emitting particle is calculated as
 \begin{equation}
  I(\xi,\phi,\nu_{e}) =
  \frac{\epsilon(\xi,\phi)}{4\pi}\frac{e^{-(\nu_{e}-\nu_{0})^{2}/{2\sigma^{2}}}}{(2\pi)^{1/2}\sigma},
 \end{equation} where $\nu_{e}$ is the emission frequency and
$\nu_{0}$ is the rest frequency corresponding to H$\alpha\ \lambda
6564.6$ and $\sigma$ is the local ``broadening parameter''
\citep{CeH89}.

\begin{figure}
\centering
\includegraphics[width=8.cm]{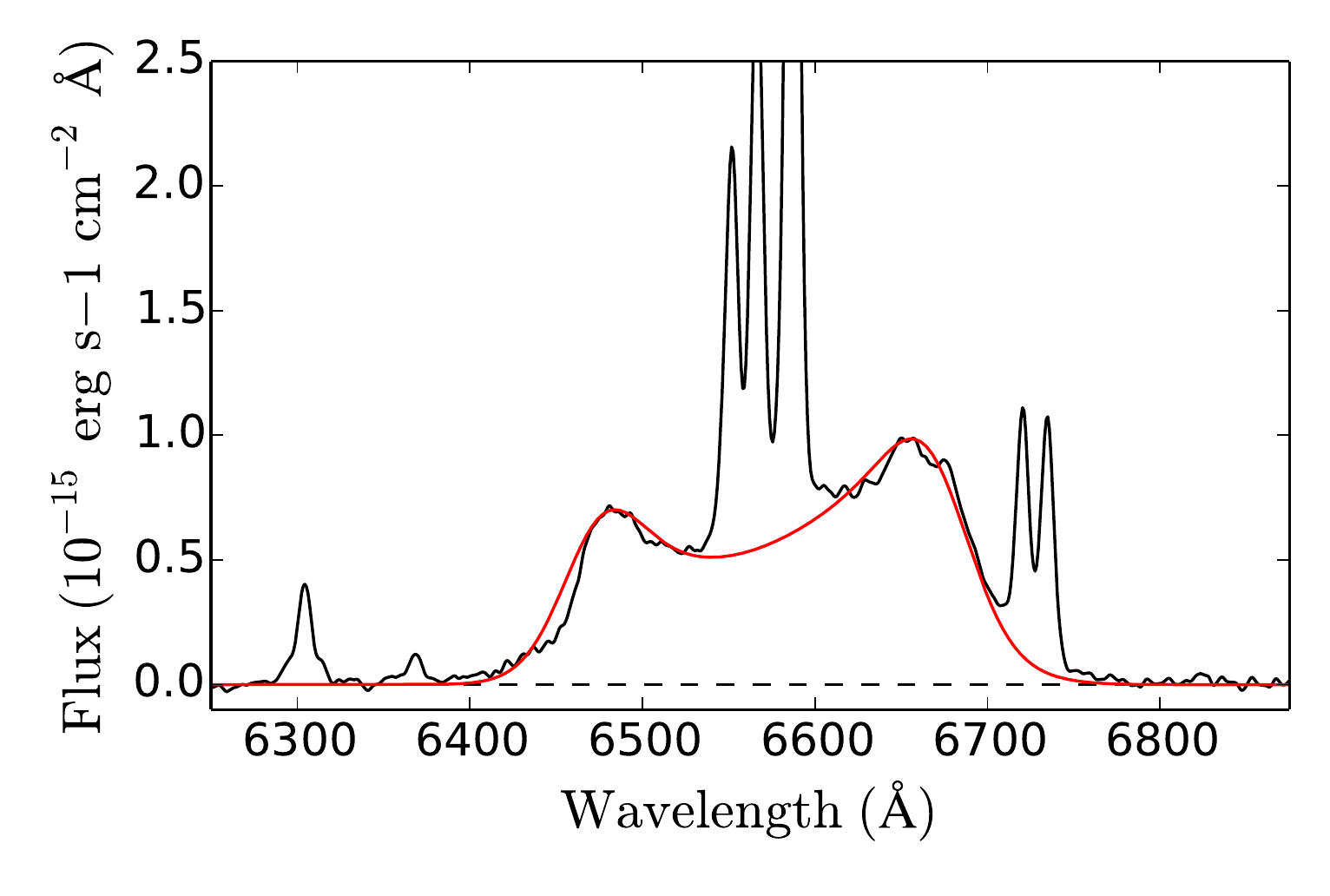}
\caption{Fit of the accretion disk model to an H$\alpha$ profile of the LLAGN in NGC\,1097
\citep{Schimoia12}. The solid black line is the observed nuclear emission spectrum
and the red solid line is the accretion disk model.}
\label{n1097}
\end{figure}

The fit of this model to the  H$\alpha$ double-peaked profile of NGC\,1097 observed on 2010 March 4  is shown
in Fig.\,\ref{n1097}. The model parameters are listed in Table \ref{n1097_table}. This model and its variants 
have been used also to 
reproduce varying double-peaked profiles of 5 broad-line radio galaxies by \citet{Gilbert99} and of 14 other AGNs from \citet{Gezari07} and \citet{Lewis10}.

\begin{deluxetable}{l c c }
\tablecolumns{8}
\tablecaption{Accretion disk parameters for H$\alpha$ profile of NGC\,1097 
and H$\beta$ RMS profile of NGC\,3227}
\tablehead{
Disk component   		&NGC\,1097      &NGC\,3227 \\
& & (RMS spectrum)}
\startdata
$\xi_1=\xi_{sp}$			&450	 	&2000		\\
$\xi_2$				&1600		&5000		\\
$\xi_q$				&1200		&3125		\\
$i$				&34	&16		\\
$\phi_0$			&140		&20		\\
$q_1$				&-2		&-2		\\
$q_2$				&3		&3		\\
$A$				&3.0		&0.6		\\
$p$				&50		&-20		\\
$\delta$($^\circ$)	&70		&40		\\
$\sigma$			&900		&480		\\
${\rm FWHM}$              &10555		&6253   	\\
shift (km\,s$^{-1}$)	&$\ldots$	&-686\\
\enddata
\tablecomments{Units of $\xi_1$, $\xi_2$ and $\xi_q$ are gravitational radii.}
\label{n1097_table}
\end{deluxetable}

\section{Double-peaked Balmer-line profiles in nearby Seyfert\,1 galaxies}

Although previous studies -- as those mentioned above -- have focused 
on fitting the broadest and cleanest double-peaked profiles, 
many nearby Seyfert\,1 nuclei also show broad Balmer  lines with double peaks or shoulders. This can be 
observed, for example, in the nuclear spectra of Seyfert 1 galaxies which have been the subject of spectroscopic monitoring campaigns, such as NGC\,4151 \citep{Maoz91, Kaspi96, Bentz06}, NGC\,5548
\citep{Netzer90, Peterson02, Bentz07, Bentz09, Bentz10, Denney10}, NGC\,3227 \citep{Winge95, 
Denney10} and NGC\,3516 \citep{Denney10}. \citet{Denney10} present the results of
reverberation mapping of other 4 Seyfert 1 nuclei besides NGC\,5548 and NGC\,3516.
With the exception of one source, either the H$\beta$ profiles are double-peaked or the
root-mean-square residual $RMS$ spectra are double-peaked.
This indicates that, even if the profile is not clearly double-peaked (e.g. due to the presence of other 
components), the most variable part of the profile is double-peaked, supporting the idea that the
innermost part of the BLR (that varies most, as it is closer to the ionizing source) is a disk.

\begin{figure}
\centering
\includegraphics[width=8cm]{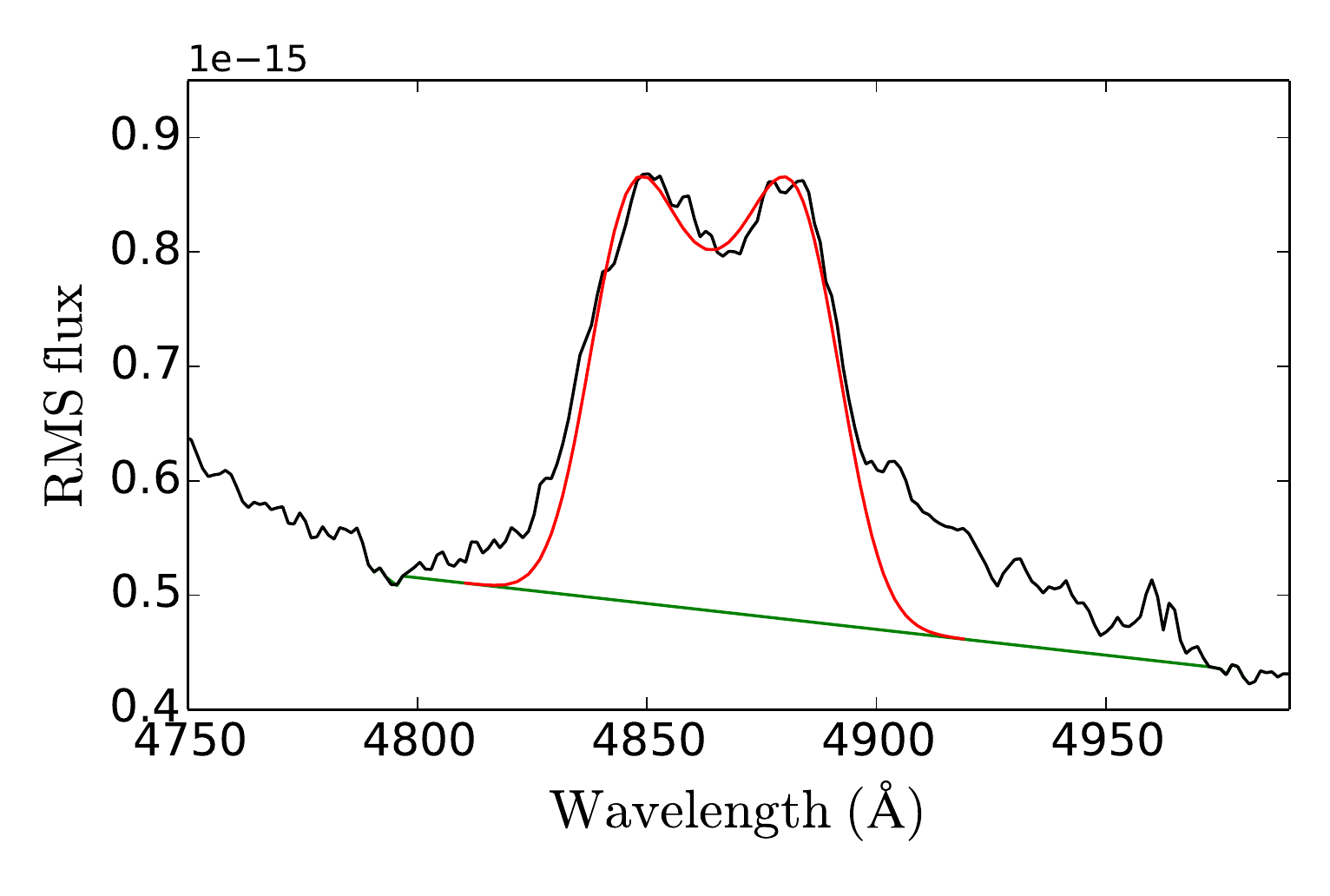}
\caption{The H$\beta$ $RMS$ spectrum of NGC\,3227 from \citet{Denney10}: the red solid line is the fitted accretion disk model profile,
while the green solid line is the adopted continuum.}
\label{n3227}
\end{figure}

We show in Fig.\,\ref{n3227} the result of our attempt to fit the $RMS$ H$\beta$ profile of  NGC\,3227
using our disk model and data from  \citet{Denney10}: although the profiles are not always double-peaked, the $RMS$ spectrum is.
The parameters of the fit are listed in Table \ref{n1097}.
Fig.\,\ref{n3227} shows that the model does fit the double-peaked component although there is, in addition,
an even broader component, revealed by the extended red wing that could not be fitted by our model. 
This component probably originates at gas at higher velocities closer to the SMBH and/or in inflow or outflow, for example.
This interpretation is consistent with the velocity-delay map of \citet{Denney12}: they conclude that there is evidence for
an overall Keplarian disk-like structure, but potential evidence for an outflow (or inflow) at large velocities, particularly on the red side 
of the line, that appears to be at larger radii than the disk-like structure found in the main line core velocities. 
Five other cases of more distant and luminous Type 1 AGN showing similar $RMS$ profiles have been reported by \citet{Grier12A,Grier12B}, 

%

%


\section{Type 1 AGN in the Palomar Spectroscopic Survey of Nearby Galaxies}

The Palomar spectroscopic survey \citep{FS85,ho95,Ho97d} -- consisting of spectra of 486
northern galaxies -- 
has a detection limit for emission lines at an equivalent width ${\rm EW} \ga
0.25$\,\AA. Of the total sample of 486 galaxies, the nuclear spectra
of 211 have been classified as AGN, with 46 among them classified as of Type 1. 



Table 1 of \citet{Ho97d} lists the properties of the 46 Type 1 nuclei of the Palomar sample: 34 are classed as showing ``definite" detection
of a broad H$\alpha$ line, while for the remaining 12, there is only a ``probable" detection of a faint
broad H$\alpha$ component. The fit of the line profiles in
these ``probable" cases shows that the broad component is not well constrained 
and may be due instead to complex kinematics of the narrow-line region (hereafter NLR).
We have inspected the H$\alpha$ profiles in the Palomar spectra of the 34 nuclei considered to present
definite detection of a broad H$\alpha$ line, and searched in the
literature for previous studies with analysis of these profiles. We list in the top part of
Table\,\ref{sy1_tab} the 12 galaxies that show clearly double-peaked or
flat-topped H$\alpha$ profiles, giving in the last column of the table the reference in which the
profile is shown to be double-peaked (or disk-like). Some of these profiles were found not in the Palomar
spectra but in nuclear spectra obtained with the STIS
spectrograph aboard the 
{\em Hubble Space Telescope (HST)}: NGC\,4203\,\citep{Shields00}, NGC\,4450\,\citep{ho00} and
NGC\,4579\,\citep{Barth01}. The higher angular resolution of the {\em HST}
spectroscopy allows the detection of even fainter lines than those detected in the Palomar survey as the
contribution of the stellar population decreases even further 
in the narrower STIS  slits ($0\farcs1-0\farcs2$).

In order to further support our argument that broad double-peaked profiles are frequently
found in Type 1 AGN, we list in the bottom part of Table\,\ref{sy1_tab} six additional Seyfert 1 galaxies
with broad double-peaked or disk-like Balmer-line profiles that are not in the
Palomar survey (most of them being southern objects), but are Type 1 AGN at distances similar to
to those of the Palomar galaxies. The double-peaked or disk-like profiles of these galaxies
were discovered serendipitously.

\begin{deluxetable}{l c c c c c}
\tablewidth{8cm}
\tablecolumns{5}
\label{sy1_tab}
\tablecaption{Nearby Type 1 AGN with Double-Peaked Balmer Profiles}
\tablehead{
Galaxy   &Activity	&$D$(Mpc)	 &Morphology &F$_{dp}$/F           &Ref.}
\startdata
{\it Palomar} &&&&&\\
NGC\,3031   &LLAGN &0.66     &SA(s)ab     &0.57         &A \\
NGC\,3227   &Sy\,2           &20.3	  &SAB(s)a pec     &0.71	          &B\\
NGC\,3516   &Sy\,1.5            &37.2    &(R)SB0$^0?(s)$     &0.91          &B\\ 
NGC\,4051   &Sy\,1            &12.7   &SAB(rs)bc     &0.74          & E\\
NGC\,4151   &Sy\,1.5      &17.0    &(R')SAB(rs)ab?    &0.63          &C\\
NGC\,4203   &LLAGN          &18.6    &SAB0$^-$?     &0.34$^*$           &D \\
NGC\,4235   &Sy\,1.2          &37.7    &SA(s)edge-on    &0.84           &E \\
NGC\,4395   &Sy\,1.8	     &8.0      &SA(s)m    &0.63           &E\\
NGC\,4450   &LLAGN         &31.1      &SA(s)ab   &0.20$^*$            &F \\
NGC\,4579   &Sy\, 2           &25.3     &SAB(rs)b     &0.21$^*$            &G\\
NGC\,5273   &Sy\,1.9         &17.8    &SA0$^0$(s)     &0.84            &E \\
NGC\,5548   &Sy\,1.5	    &73.4      &(R')SA0/a(s)  &0.96            &E\\
\hline\\
{\it Southern} &&&&&\\
NGC\,1097 &LLAGN	    &15.1     &SB(s)b   &\ldots		&H\\
NGC\,3065 &LLAGN            &28.3    &SA0$^0$(r)    &\ldots         &I\\
IC\,4329A	 &Sy\,1.2	    &69.6	  &SA0$^+$? edge-on  &\ldots           &J\\
NGC\,2617 &Sy\,1.8	    &62.3	  &Sc  &\ldots    	&K\\
NGC\,3783 &Sy1.5        &44.3      &(R')SB(r)ab  &\ldots            &L \\
NGC\,7213  &LLAGN    &21.1      &SA(s)a?  &\ldots           &M\\
\enddata
\label{sy1_tab}
\tablecomments{
Column (1): Galaxy identification; (2): Activity type; (3) distance (NED, 3K CMD); 
(4): galaxy morphology; (5): fractional contribution of broad  
H$\alpha$ to the total H$\alpha$+[N\,{\sc ii}] emission; $^*$ means that the double-peaked profile was
identified only with {\em HST} STIS spectra; (6):
reference for the double-peaked profile: (A)\citet{Bower96}, (B)\,\citet{Denney10}, 
(C)\,\citet{Bentz06}, (D)\,\citet{Shields00}, 
(E)\,\citet{Ho97d}, (F)\,\citet{ho00}, (G)\,\citet{Barth01}, (H)\,\citet{SB93}, 
(I)\,\citet{Eracleous01}, (J)\,\citet{Winge}, (K)\,\citet{Shapee14}, (L)\,\citet{Stirpe94}
(M)\,\citet{Filippenko84}.}
\end{deluxetable}

\subsection{Fit of the double-peaked profiles of Type 1 AGN in the Palomar sample}

In order to further demonstrate that the profiles that appear double-peaked in the Palomar spectra
can be attributed to the outer parts of an accretion disk, we now use
our model to fit the broad H$\alpha$ profile in these spectra \citep{Ho97d} for the galaxies
of Table\,\ref{sy1_tab} that showed high enough signal-to-noise ratio in the broad profiles to constrain the fit:
NGC\,3516, NGC\,4151, NGC\,4235, NGC\,5273 and NGC\,5548.

\subsection{Subtraction of the stellar component}
\begin{figure}
\centering
\includegraphics[width=8cm]{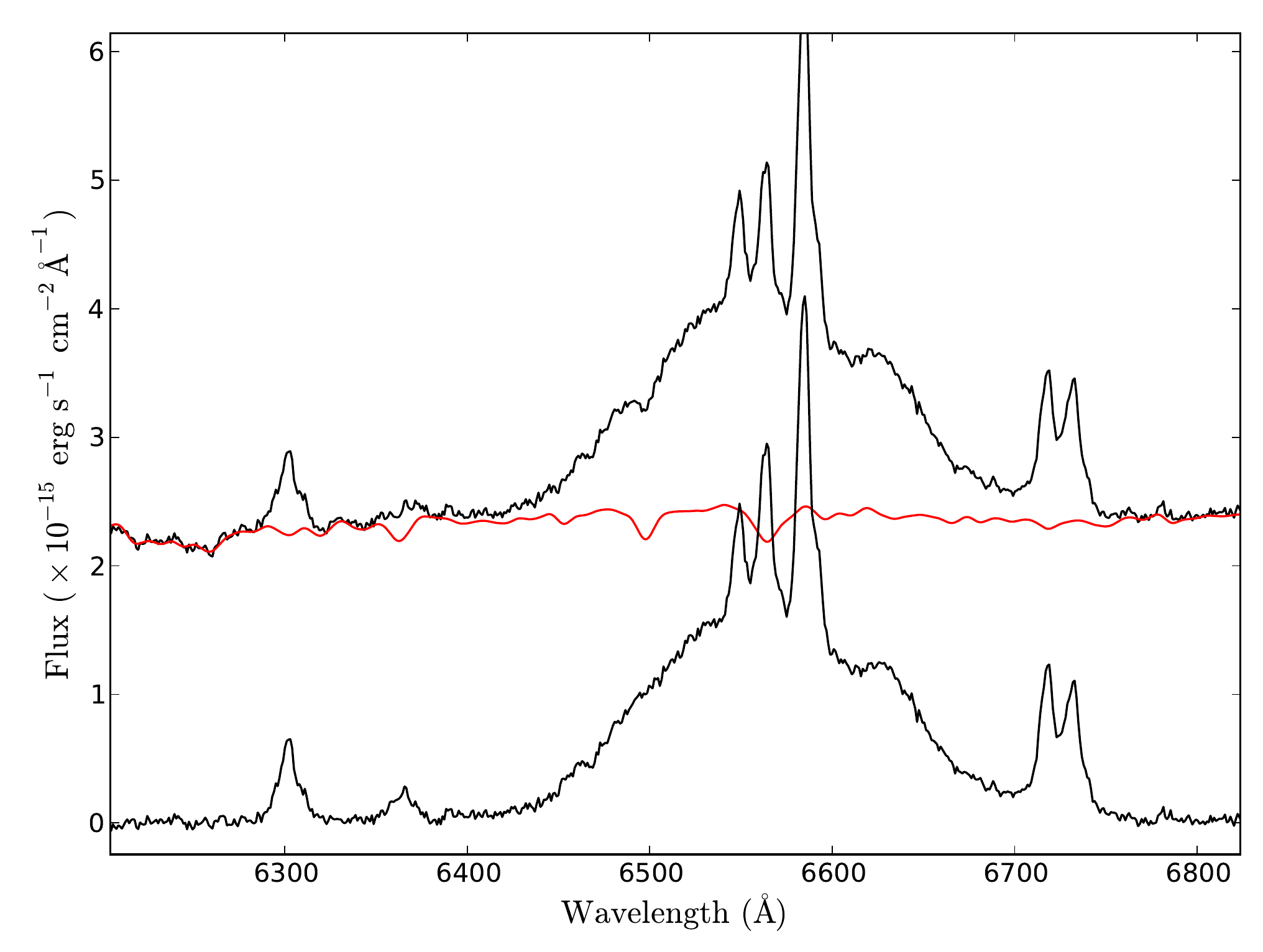}
\caption{Subtraction of the stellar population template NGC\,4235. Top: Palomar
spectrum in black; stellar population template (NGC\,4371, from \citet{Ho97c}) in red. 
Bottom: residual nuclear emission spectrum.}
\label{pop_sub}
\end{figure}

\begin{figure}
\centering
\includegraphics[width=8cm]{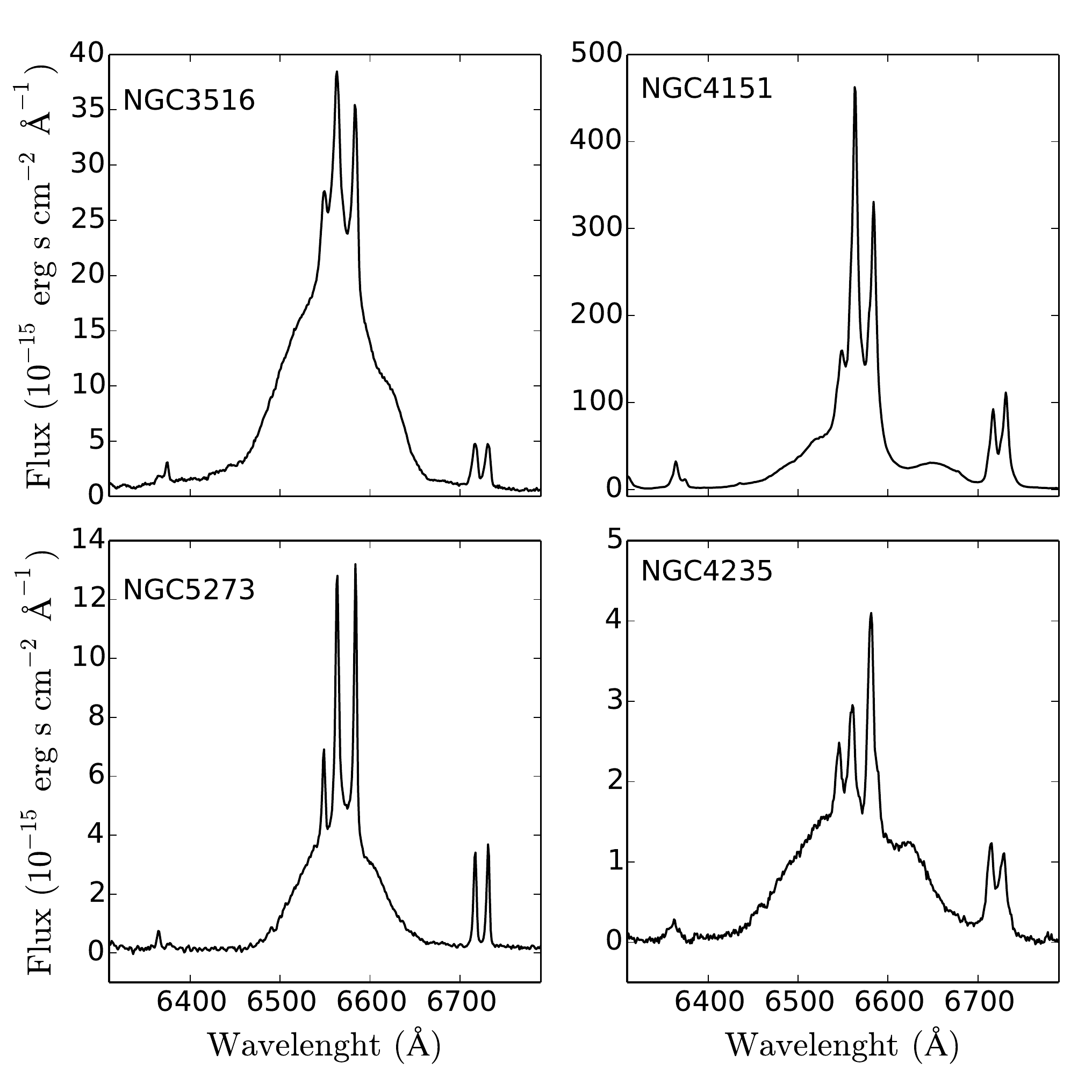}
\caption{Nuclear emission spectra of NGC\,3516, NGC\,4115, NGC\,4235 and NGC\,5273 after the
subtraction of the contribution of the stellar population.}
\label{gals}
\end{figure}

Most of the spectra of the above galaxies show strong absorption features due the contribution 
of an underlying stellar population, with the exception of NGC\,5548, whose continuum seems
to be dominated by the AGN in this spectrum. In order to  isolate the gas emission in the other cases, 
we corrected their nuclear spectra by subtracting a stellar population template, 
using for this the Palomar spectra of the non-active galaxies \citep{Ho97c} that do not show any emission lines.
For each active galaxy, we searched for a stellar population template that reproduced its nuclear spectrum well (in the regions where there are no emission lines) and then subtracted a scaled and Gaussian convolved version of the chosen template. We used a continuum wavelength window from 6200\AA\ to 6250\AA\ to determine the scale factor between the nuclear spectrum and the stellar template. This method is very similar to
that of \citet{Ho97c} and the result is illustrated in Figure \ref{pop_sub} for NGC\,4235. 
The spectra of the other four galaxies after the subtraction of the stellar component are shown in Figure \ref{gals}.

\subsection{Fit of the disk component}


We have used the ``Saturated Spiral Model''  \citep{Schimoia12} that has provided the best fits in our previous studies of the H$\alpha$ double-peaked profile of NGC\,1097 and in an on-going study of a series of nuclear spectra of another LINER/Seyfert 1 galaxy, NGC\,7213 \citep{Schimoia17}. We constrained the double-peaked component by requiring it to fit the visible peaks and/or shoulders of the profile, and at the same time trying to fit the wings. In order to reduce the number of free parameters, we fixed the index $q_2$ of the outer part of the emissivity law ( $\xi_q<\xi<\xi_2$) at $q_2=3$, motivated by the fact that for an accretion disk ionized/illuminated by an external source above the disk (as usually adopted as the source of ionization for the gas in the disk) the radial dependence of the surface emissivity is $\epsilon(\xi)\propto\xi^{-3}$ \citep{Dumont90}. Only in the case of NGC\,4235 we adopted $q_2=2$, as this allowed an improvement in the fit. We also fixed the inner radius where the spiral arm begins at $\xi_{sp}=\xi_{1}$, that worked well except for NGC\,4151 (see discussion in Sec. \ref{results}).

In order to try to fit the broad wings of the profiles, it was necessary to increase the emissivity of the inner part of the disk (for $\xi_1<\xi<\xi_q$) relative to our previous modelling (e.g. of the double-peaked profile of NGC\,1097 and NGC\,7213), adopting and index $q_1=0$. This index implies a constant emissivity for the region within the radius $\xi_q$, while in our previous studies of NGC\,1097 and other LINERs the best fit required an emissivity increasing between $\xi_1$ and $\xi_q$ (with $q_1=-2$).

\subsubsection{An aditional broad component}

\citet{Schimoia17} monitored the double-peaked profile of the LINER/Seyfert 1 nucleus of NGC\,7213 and find that the $RMS$ spectrum shows, besides the double-peaked component, a central ``hump" in the profile well reproduced by a broad (FWHM $\approx\,2200$\,km\,s$^{-1}$) Gaussian component with centroid velocity close to systemic. We have interpreted this finding as due to the contribution of BLR clouds farther away from the ionizing source than the disk, having thus lower orbital velocity. We find this to be the case also for most Seyfert 1 profiles fitted here: besides the disk component, a broad Gaussian component with centroid velocity close to systemic is also necessary to fit the H$\alpha$ profile.

\begin{figure*}
\centering
\vspace{-3cm}
\includegraphics[width=17cm, angle=90, scale=0.7]{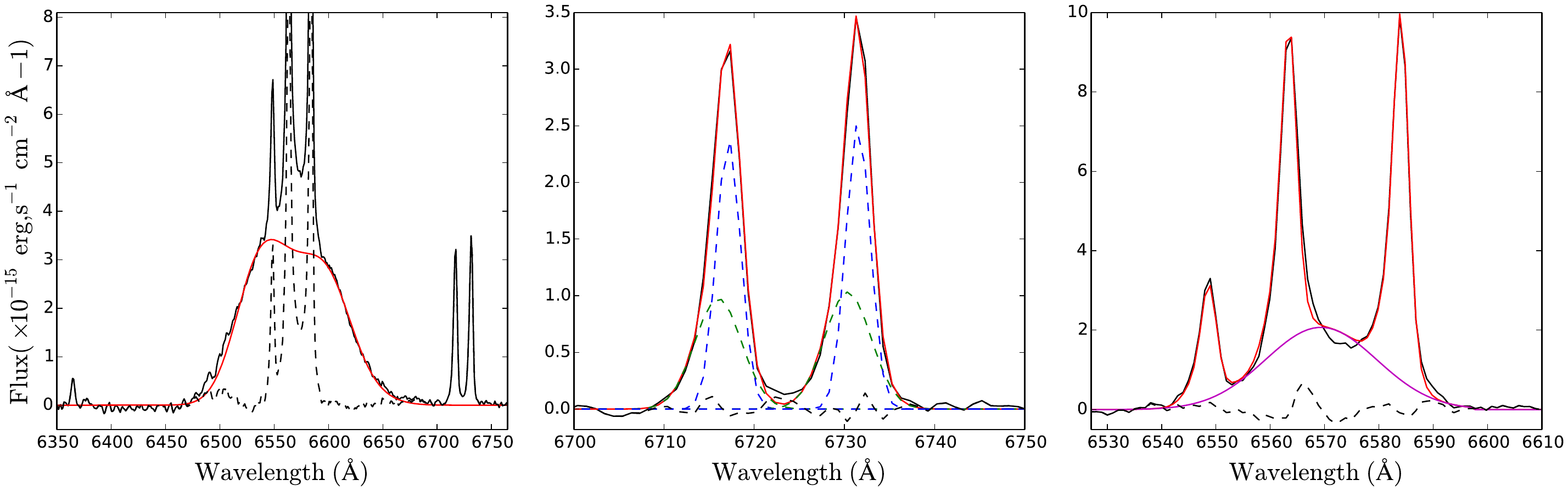}
\vspace{-3.5cm}
\caption{The aditional broad component in NGC\,5273.
\emph{Left}: observed profile in black continuous line, fitted double-peaked component in red, residuals in dashed lines;  \emph{Center}: fit of the [S\,II]$\lambda\lambda$6717,31 lines  with two Gaussian components, represented in green and blue; \emph{Right:} fit of the [N\,II]$\lambda\lambda$6548,84 and narrow H$\alpha$ components constrained to the same Gaussian components fitted to the [S\,II] lines (in velocity space) plus the additional broad component needed to fit the H$\alpha$ line. }
\label{fitnarrow}
\end{figure*}

In order to constrain this additional component, we proceeded as follows. We fitted the profiles of each of the  [S\,II]$\lambda\lambda$6717,31 lines using two Gaussian curves, necessary (and sufficient) in order to fit the profiles including eventual wings. Each Gaussian fitted to the $\lambda$6731 line was constrained, in velocity space, to have the same width, relative intensity and relative central velocity as each corresponding Gaussian fitted to the $\lambda$6717 line. Then, under the assumption that the narrow components of H$\alpha$ and [N\,II] lines originate from the same region as the [S\,II] lines, we used the [S\,II] profiles in velocity space to fit the narrow components of the H$\alpha$ and [N\,II]$\lambda\lambda$6548,84 lines -- the last two also constrained to have an intensity ratio of 3 \citep{Ost06}. The result was that, besides the disk and the constrained narrow components, an additional component -- well represented by a broad (1000\,km\,s$^{-1}\le$FWHM$\le$2000\,km\,s$^{-1}$) Gaussian -- was necessary to fit the residual H$\alpha$ flux in most cases (with the exception of NGC\,4273). This procedure is illustrated in Fig.\,\ref{fitnarrow}  for the case of NGC\,5273.

The additional broad component was fitted by the following Gaussian curve:

\begin{equation}
F_{G}(\lambda)= \frac{A_G}{\sqrt{2 \pi {\sigma_G}^2}} \exp{ \left[ - \frac{(\lambda - \lambda_0 +
\Delta\lambda)^{2}}{2{\sigma_G}^2} \right]}
\end{equation}

\noindent where $\lambda_0$ is the rest wavelength of  H$\alpha$, $\Delta\lambda$ is t
displacement of the Gaussian component relative to the H$\alpha$ narrow component,
$\sigma_G$ is the standard deviation and $A_G$ is the amplitude of the Gaussian component. 





Following \citet{Popovic04}, we have calculated the relative flux contribution between the Gaussian and disk components as:

\begin{equation}
Q=\frac{\int_{-\infty}^{+\infty} F_G(\lambda) d\lambda}{\int_{-\infty}^{+\infty} F_D(\lambda)
d\lambda}
\end{equation}

\noindent where $F_D(\lambda)$ is the flux of the disk component.

\subsection{Results from the fits}
\label{results}

The parameters derived from the fits of the disk component as well as the FWHM, velocity shift and  integrated fluxes are listed in Table \ref{two_components}, where the uncertainties in the parameters were obtained by varying each parameter separately around the best fit value while keeping all the others parameters fixed. In the bottom part of the Table we list the integrated fluxes, FWHM and velocity shift of the Gaussian component, while in the last line we list the value of the $Q$ parameter. In the last column of Table \ref{two_components} we list the average values of the disk parameters for all the galaxies and $rms$ deviation from the mean for all the listed properties, including in this average the values for NGC\,1097, listed in Table \ref{n1097_table}.

\begin{figure*}
\centering
\includegraphics[width=17cm]{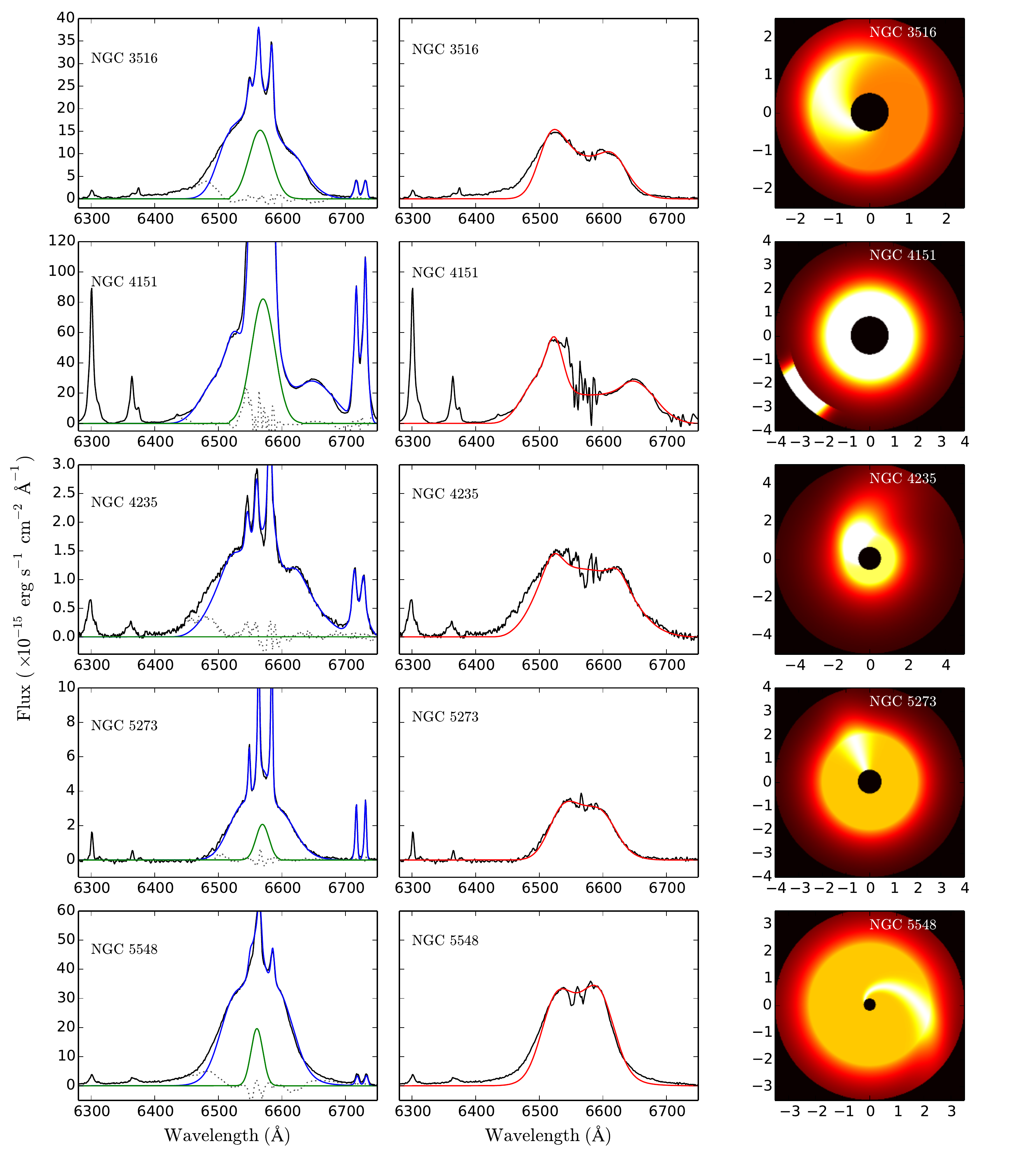}
\caption{Fits of the disk and broad Gaussian component to the H$\alpha$ profiles together with those of the narrow components of H$\alpha+$[NII] and [SII] for NGC\,3516, NGC\,4151, NGC\,4235, NGC\,5273 and NGC\,5548.
\emph{Left column}: observed spectrum in continuous black lines; broad Gaussian component in green and combined fit of the disk, broad Gaussian and narrow components in  blue; the dotted line shows the residuals between the observed profile and fit;
\emph{Central column}: spectrum after subtraction of the H$\alpha$ broad Gaussian and narrow components above in black; best disk model in red; \emph{Right column}: emissivity map of the disk component (yellow-white corresponding to the maximum emissivity); disk is seen face-on; when inclined, observer should be at the bottom of the figure, with the line-of-nodes running along the horizontal direction. Axes labels are in units of 1000$\xi$.}
\label{gals_models}
\end{figure*}

Figure \ref{gals_models} shows, in the left column, the fits of the disk and broad Gaussian component to the H$\alpha$ line, together with those of the narrow components to the H$\alpha+$[N\,II] and [S\,II] line profiles. In the central column we show only the disk component overplotted on the emission-line profiles after the subtraction of the broad Gaussian component and narrow components. In all figures, the profiles of the narrow components are well reproduced by the [S\,II] line profiles, as illustrated for NGC\,5273 in Figure \ref{fitnarrow}. The parameters of the disk component, whose profiles are shown in the central column, were used to generate the cartoon shown in the last column of Fig.\,\ref{gals_models} with a rendering of the disk (seen face-on) using the best fit parameters listed in Table \ref{two_components}.


\subsubsection{The Disk component}

Table \ref{two_components}, combined with the results of the fit for NGC\,1097 in Table \ref{n1097_table}  shows that the inner radius of the disk ranges  between 500 and 800 $R_g$, the outer radius between 2500 and 5500 $R_g$, and the saturation radius between 1300 and 2600 $R_g$. Average values are listed in the last column of Table\,\ref{two_components} (we have included in this calculation the values for NGC\,1097), showing $rms$ variations of $\sim$\,30\%. 

The emissivity distribution of the disk is approximately the same for all Seyfert 1 galaxies: it is constant from the inner radius of the disk up to the saturation radius (emissivity law index $q_1=0$) and then decreases with emissivity law index of $q_2=3$ for all cases but for NGC\,4235, whose profile is better fitted with $q_2=2$. We note that we have obtained a better fit with $q_1=0$ for the Seyfert 1 galaxies than with $q_1=-2$ (this worked better for NGC\,1097 and NGC\,7213). This may indicate a difference in the geometry/structure of the ionizing source between the LLAGNs and Seyfert 1 galaxies, such that in the latter the ionizing source shows a larger incidence of ionizing radiation in the inner part of the disk than in the LLAGNs, where the incidence or ionizing radiation seems to increase from the inner radius up to the saturation radius.

The spiral arm has a small contrast (between 1 and 2.5) and begins at the inner radius of the disk, with the exception of NGC\,4151, in which case we had to make the spiral to begin close to the outer border of the disk and give it a much larger contrast (88) in order to reproduce a strong blue ``hump" in the profile. The parameters characterizing the arm in this case are such that the resulting structure can be better described as a ``hot spot" in the outer border of the disk.

The inclination angle of the disk (relative to the plane of the sky) is in the range
$17^{\circ}<i<38^{\circ}$ (including NGC\,1097), meaning that we are seeing the disk closer to face-on than edge-on, in agreement with what is expected for Type 1 AGN in the simple Unified Model \citep{Antonucci93}.

\subsubsection{The case of NGC\,5548}

The broad H$\beta$ profiles of NGC\,5548 have been modelled by \citet{Pancoast14b} as originating in a BLR composed by clouds with a flattened distribution between an inner an outer radius, thus in approximate agreement with a disk model like ours. Nevertheless, these authors derive an inclination $i$ for the flat distribution of clouds of $i=39\pm11^\circ$, while we have obtained a much smaller inclination $i=19\pm2^\circ$. 

On the other hand, the minimum radius of 1.39 light days (l.d.) and mean radius of 3.31 l.d. obtained by \citet{Pancoast14b} correspond to 809 and 1927 gravitational radii ($R_g$), respectively, for their quoted BH mass of 3.23$\times$10$^{7}$M$_\odot$, consistent with our estimate for the inner radius of the line emitting disk of $\xi_1=600\pm300$ ($R_g$) and the saturation radius of 2600$\pm$500\,$R_g$ (see Table\,\ref{two_components}). We also note that, if the mean radius of the disk is a characteristic radius from which most of the line emission origintes, as is the case of the saturation radius in our modeling, and we calculate the Keplerian velocity $v_K$ at 3.31 l.d. from the BH, we obtain $v_K=7086$\,km\,s$^{-1}$. If the inclination is $i=19^\circ$ the observed velocity should be $v_K \sin(i)=2307$\,km\,s$^{-1}$, in approximate agreement with our measured ${\rm FWHM}=$5754\,km\,s$^{-1}$ (considering that the velocity is approximately half the FWHM value). For $i=39^\circ$, $v_K \sin(i)=4459$\,km\,s$^{-1}$, that would imply ${\rm FWHM}\ge$9000\,km\,s$^{-1}$, much broader than the value we have measured. 


\subsubsection{The Broad Gaussian component}

The broad Gaussians have full-width at half maximum in the range $1000\le {\rm FWHM} \le 2000$\,km\,s$^{-1}$, and small velocity shifts relative to the centroid velocity of the narrow-line components, between $-39$\,km\,s$^{-1} <\Delta v <+$170\,km\,s$^{-1}$.
We attribute these components to more distant clouds from the SMBH than the disk, thus having lower velocities. 

The parameter $Q$ that measures the contribution of the Gaussian component relative to the disk component ranges between 0.11 and 0.67, with the Gaussian being strongest for NGC\,4151 ($Q=0.67$) and NGC\,3516 ($Q=0.47$).

The presence of double-peaked disk components in nearby Type 1 AGNs is also supported by reverberation mapping studies of the H$\beta$ broad line profile.
\cite{Denney10} find that 5 of the 6 Seyfert 1 nuclei they studied (which include NGC\,3516 and NGC\,5548) show double-peaked or flat-topped broad H$\beta$ profiles in either the mean or $RMS$ spectrum (or both). The same is found for other 5 more distant and luminous Type\,1 AGN by \citet{Grier12A,Grier12B}. \citet{Peterson04} show that the $RMS$ spectrum of NGC\,4151 is also double-peaked, while \citet{Doroshenko12} show the same also for Mrk\,6.

\begin{deluxetable*}{l c c c c c c}
\tablecolumns{7}
\tablecaption{Parameters of the disk plus broad Gaussian modelling}
\tablehead{
Component   &NGC\,3516	                            &NGC\,4151                        &NGC\,4235                       &NGC\,5273                        &NGC\,5548                        &Average$\pm$RMS}
\startdata

{Disk} &&&&&&\\
$\xi_1$		&500$^{+300}_{-100}$	&800	$^{+200}_{-300}$        &600$^{+200}_{-300}$	&500$^{+200}_{-400}$	&600$^{+300}_{-300}$  	&575$\pm$125\\
$\xi_2$		&2500$^{+300}_{-300}$	&4000$^{+100}_{-100}$	&5000$^{+1500}_{-500}$	&4000$^{+1500}_{-1500}$	&3500$^{+800}_{-800}$ 	&3433$\pm$1211 \\
$\xi_q$		&1500$^{+400}_{-200}$	&1550$^{+150}_{-100}$	&1300$^{+300}_{-300}$	&2000$^{+400}_{-300}$	&2600$^{+500}_{-500}$ 	&1692$\pm$524 \\
$i$			&20$^{+2}_{-2}$			&38$^{+2}_{-2}$		         &31$^{+3}_{-3}$		         &17$^{+2}_{-2}$		         &19$^{+2}_{-2}$                  &27$\pm$9\\
$\phi_0$		&240$^{+20}_{-30}$		&310$^{+5}_{-5}$		&160$^{+5}_{-5}$		&210$^{+30}_{-30}$	         &45$^{+40}_{-25}$	 	&$\dots$\\
$q_1$		&0		                           &0     	                           &0		                           &0		                           &0		                           &-0.33$\pm$0.82\\
$q_2$		&3		                           &3		                           &2		                           &3		                           &3		                           &2.8$\pm$0.4\\
$A$			&1.6$^{+0.2}_{-0.2}$		&88$^{+5}_{-5}$   	         &2.5$^{+1.5}_{-0.5}$		&1.0$^{+0.4}_{-0.4}$		&1.5$^{+0.2}_{-0.2}$		&$\dots$ \\
$p$			&-20$^{+5}_{-30}$		&90$^{+60}_{-60}$   	         &-30$^{+6}_{-4}$	        &90$^{+50}_{-50}$		&-20$^{+15}_{-50}$		&27$\pm$57  \\
$\delta$($^\circ$)	&90$^{+40}_{-20}$	&20$^{+2}_{-2}$   	         &60$^{+30}_{-30}$	        &30$^{+20}_{-20}$		&30$^{+10}_{-10}$ 		&50$\pm$28 \\
$\sigma$		&750	$^{+50}_{-50}$	         &500$^{+50}_{-50}$		&500$^{+100}_{-100}$       &850$^{+150}_{-100}$		&900$^{+50}_{-50}$ 	        &733$\pm$189 \\
${\rm FWHM (km\,s^{-1}}$&6932$^{+680}_{-680}$ &8959$^{+685}_{-1052}$	 &7986$^{+454}_{-272}$  &5374$^{+228}_{-273}$	   &5754$^{+134}_{-101}$  &7001$\pm$ 1501\\
$shift$ (km\,s$^{-1}$)	&0	                           &457                                    &229 	                         &0  	                                   &-457                                    &38$\pm$303 \\
$Flux$		        &1434	                  &5636	                           &174		                &234                                     &3783                                    &2252$\pm$2390 \\
\hline
&&&&&&\\
{Broad Gaussian} &&&&&&\\
${\rm FWHM (km\,s^{-1})}$        &1919 	&1981	&$\ldots$	   &1101  &885 &1472$\pm$560\\
$shift$ (km\,s$^{-1}$) &68    &171    &$\ldots$     &149  &-39 &87$\pm$95 \\
$Flux$                                         &678       &3785          &$\ldots$     &53     &416  &1233$\pm$1720\\
\hline
&&&&&&\\
$Q$				&0.47		&0.67		 &$\ldots$    &0.23       &0.11   &0.37$\pm$0.25 \\

\enddata
\tablecomments{Units of $\xi_1$, $\xi_2$ and $\xi_q$ are gravitational radii. Fluxes units are 10$^{-15}$ erg\,cm$^{-2}$\,s$^{-1}$. The orientation of the spiral arm $\phi_0$ is measured counterclockwise from the bottom of Fig.\,\ref{gals_models} at the inner radius where it begins (with exception of NGC\,4151). Uncertainties in fluxes and FWHM values are smaller than 10\%. The calculated average values (plus the standard deviation RMS) listed in column (7) include the fitting parameters for NGC\,1097 listed in Table \ref{n1097_table}.}
\label{two_components}
\end{deluxetable*}

\section{The statistics of Double-Peaked Profile Emitters among Type 1 nuclei}

Type 1 AGN in the Palomar Survey comprise between 34 (with ``unambiguous" broad lines) and 46
(including broad lines that could be due to multiple components in the narrow lines) of the 211 AGN, thus
$\sim$\,20\% \citep{Ho97d,ho08}. We will use this
percentage to put approximate constraints on the half-opening angle $\theta_m$ of the 
obscuring torus in the simple approximation of the Unified Model scenario
in which the geometry of the escaping radiation is that of a cone 
\citep{Antonucci93}. We now know that the ``torus'' is most
likely a distribution of dusty clouds, and that its properties vary with the AGN luminosity
\citep[e.g.,][]{Elitzur12,Ichikawa15}, but the simple
Unified Model can give us some insight on relative probabilities.
In the simple model, the 20\% fraction of Type 1 AGN implies that the ``free" solid angle of the 
AGN is $2\pi[1-\cos(\theta_m)] \approx$ 20\% of $2\pi$,  with 80\% of $2\pi$ being the solid
angle covered by the obscuring torus. Thus, in order to see directly the
AGN --- the accretion disk and BLR --- our line-of-sight has to make an angle of at most $\theta_m\approx37^\circ$
relative to the AGN symmetry axis.

\begin{figure*}
\centering
\includegraphics[width=17cm]{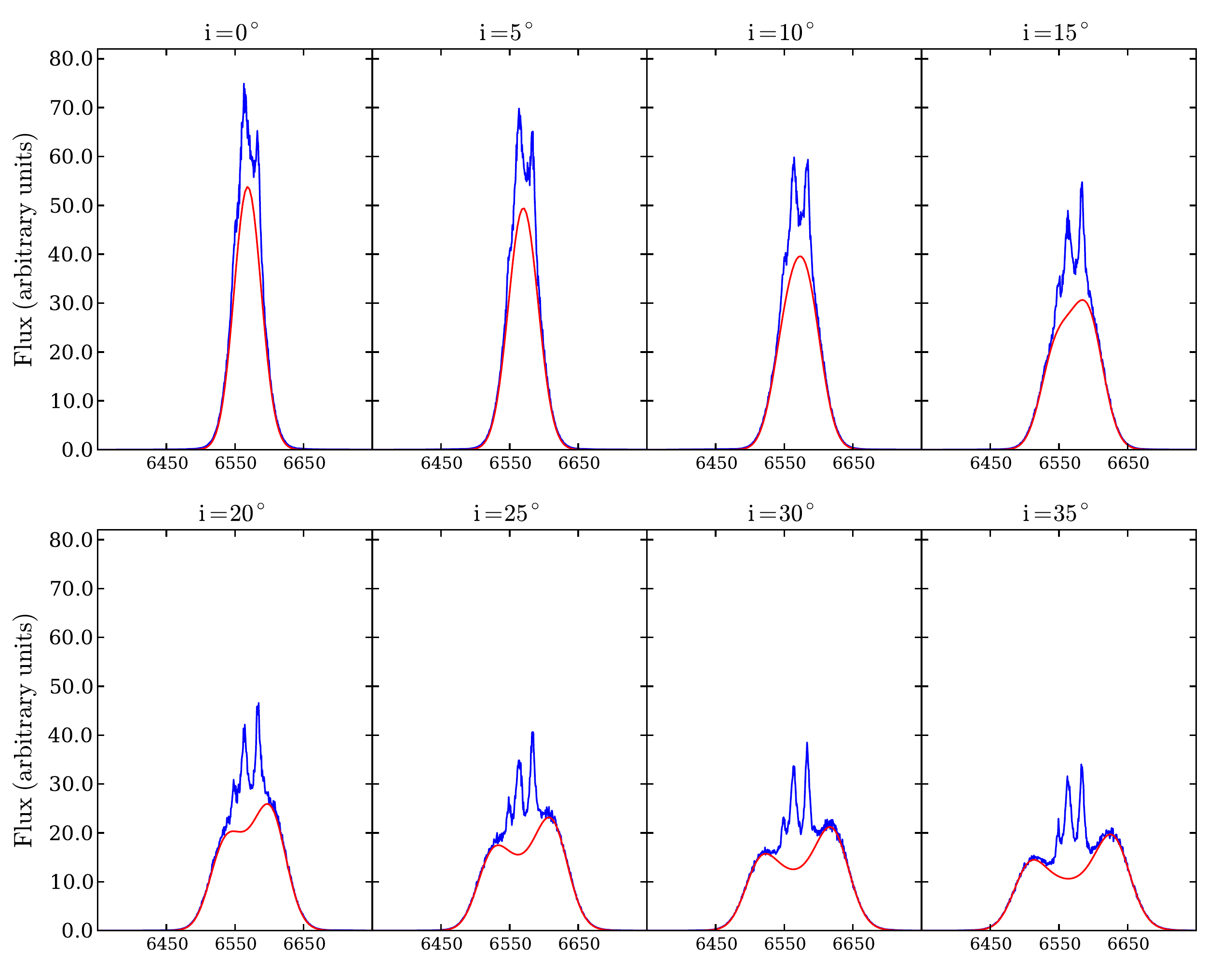}
\caption{Disk model for the H$\alpha$ profile as a function of inclination of the disk relative to the plane of the sky.
Typical narrow H$\alpha$+[NII] emission-line profiles have been superimposed on the broad double-peaked
profiles.}
\label{dp_profiles}
\end{figure*}

In order to illustrate how the double-peaked profiles would appear as a function of the orientation of
our line of sight relative to the AGN, we show
in Fig.\,\ref{dp_profiles} a series of double-peaked profile models for inclinations ranging from 0 to
35$^\circ$. The parameters of the model are the average values in Table \ref{two_components} we have obtained from the fitted models to the Palomar spectra 
applied to the H$\alpha$ profile, which is the most prominent line in the optical spectra of
AGN (as it is $\ge$\,3 times as strong as H$\beta$). In this figure, we have superposed the narrow
components of H$\alpha$ and
[N\,{\sc ii}]\,$\lambda\lambda6548,$ 6584 on the broad double-peaked profile. These lines were simulated
by Lorentzians with ${\rm FWHM} = 300$\,km\,s$^{-1}$.

Figure \ref{dp_profiles} shows that, in order for H$\alpha$ to appear double-peaked, our
line-of-sight has to make an angle of at least $\sim\,20^\circ$ relative
to the ionization axis of AGN. This means that the disk inclination 
should be between $\sim20^\circ$ and $\sim37^\circ$
for the profile to be observed as double-peaked. The inclination angles derived from our fits
are between 17$^\circ$ and 38$^\circ$, thus in good agreement with the predicted range. 
Previous fits of similar accretion disk models to double-peaked profiles, such as those observed in radio galaxies
\citep{CeH89,EH94,Lewis10}, and nearby LLAGN \citep{ho00,SB03} also show a similar range of inclinations.
Considering the inclination values listed  in Table\,\ref{two_components} 
for the 5 targets we have fitted with our disk model plus the fit to NGC\,1097, we obtain a mean value for the inclination of the disk of $i=27^\circ \pm 9^\circ$.


Assuming again the conical geometry of the simple model, the solid angle for inclinations between $20^\circ$ and $35^\circ$ is
12\% of 2$\pi$. When we compare this number
with the 20\% fraction above for type 1 AGN, we conclude that $\sim60$\% of type 1 AGN should have
double-peaked Balmer-line profiles. According to these predictions, if the total sample of Type 1 nuclei in the
Palomar sample is 34 galaxies, we should expect to find
$\sim20$ galaxies with double-peaked profiles. We find 12, thus $\sim35$\% instead of the predicted 60\%. On the other hand, the number we have found may be a lower limit, as, in some cases the double-peaked nature of the line profiles could be observed only
in very high-quality {\em HST} STIS spectra in
which the contrast between the emission lines and the host-galaxy starlight is much higher than in the
Palomar spectra. In addition, in the small apperture of the STIS spectra, the NLR contribution can also be minimized,
increasing the contrast between the BLR and NLR emission. As not all the Type 1 AGN of the
Palomar sample have been observed with {\em HST} STIS, this fraction of 35\% double-peakers may be a
lower limit. 

Another consideration is that, in the Palomar spectra, a critical property of the broad line to allow
its clear detection is its flux compared with the narrow lines.
Inspection of the fits in Figs. 4--14 of \citet{Ho97d} shows that clear broad H$\alpha$ components
can only be detected when
the fraction of the flux in the broad line is larger than 50\% of the total flux in the broad plus
narrow H$\alpha$+[N\,{\sc ii}] lines. This fraction,
obtained from \citet{Ho97d}, is listed in the fifth column of Table\,\ref{sy1_tab}. The total number
of these sources from \citet{Ho97d} is 14, with 9 (64\%)
of them having double-peaked profiles, now consistent with the $\sim$\,60\% prediction above. Another
possibility is that the ratio of the outer to inner radius of the disk is $>>10$, what would make the line profiles
singly-peaked \citep[see Figure~9 of ][]{Eracleous03}. Finally, if the accretion rate is high,
we would expect a contribution from non-disk clouds in the form of a single broad Gaussian component, as discussed
in the previous section. This component would ``fill" the double-peaked ``dip" in the profile, making it difficult to distinguish 
the double-peaked component in the total emission-line profile.

\section{Discussion}

As mentioned in Section 1, the fact that broad-line profiles of Seyfert 1 and radio galaxies seem to contain a flattened component has been noted in previous studies, including also those by \citet{WB86}, \citet{CeH89}, \citet{JB91}, \citet{Rokaki92}, \citet{Brotherton96}, \citet{Marziani96} and \citet{WeP96}.
More recently, \citet{Pancoast14a, Pancoast14b} have performed detailed dynamical modeling of the BLR
of five Seyfert\,1 nuclei. 
The main conclusion of \citet{Pancoast14b} is that the BLR geometry is  a thick disk viewed close
to face-on, in agreement with our findings. They argue also for the presence of inflows that could be due to the motion of BLR clouds in elliptical orbits around the SMBH \citep{Elitzur14}. Another possibility is that the Gaussian component represents perturbations in the surface brightness of the disk that are not captured by the single spiral arm we have adopted.




On the basis of this and previous studies we suggest that the disk is present in most BLRs, and at the higher 
Eddington ratios of Seyfert 1 galaxies, we observe additional line-emitting components corresponding to
non-disk clouds, usually at lower velocities. Some of these clouds
may be inflowing \citep{Grier13A}, just orbiting the SMBH in elliptical orbits as proposed by
\citet{Pancoast14b}, or may be part of an outflowing disk wind \citep{Elitzur14}.

\subsection{Implications for the derivation of Black Hole Masses}

The technique of reverberation mapping has provided valuable tools for the estimation of the mass of SMBHs in
AGN up to very large distances in the Universe,
solely on the basis of the observed continuum luminosity and width of the broad emission lines. The
optical luminosity allows the estimate of the
BLR radius \citep{WPM99,Kaspi05,Bentz13}, while the width of the broad profile provides the velocity
dispersion of the BLR clouds $\Delta V$.
Assuming that the motion of the BLR gas is dominated by gravity and that radiation pressure can be
neglected, the mass of the central black hole is given by \citep{Peterson04}

\begin{equation}
\label{eq:mass}
M_{\rm BH} = f \frac{R_{\rm BLR} \Delta V^2}{G}\mathrm{,}
\end{equation}

\noindent where $R_{\rm BLR}$ is the radius of the BLR, $G$ is the gravitational constant and $f$ is a
dimensionless numerical factor that depends on the structure,
kinematics and orientation of the BLR. A recent empirical determination of the average scale
factor  
is $\langle f \rangle = 4.31 \pm 1.05$ \citep{Grier13B}, based on the assumption that the relationship
between black hole mass and bulge velocity
dispersion is the same for quiescent and active galaxies. The value of $f$ is of fundamental importance
in using Equation \ref{eq:mass}
for the determination of the masses of SMBH's across the Universe, and incorporates the uncertainties
due to the geometry and inclination of the BLR.

The recognition that the disk component may be ubiquitous in Type 1 AGN has important
implications for the value of $f$ in the cases dominated by the disk component. If we consider
that the geometry of the region that varies on the shortest timescale probed by the profile variation is
a flat disk,  $f=1/\sin^2 i$, where $i$ is the inclination of the disk relative to the plane of the
sky, as the observed width of the lines $\Delta V$ is simply the rotational speed projected into the line-of-sight. In particular,
for $\langle f \rangle = 4.3$, $\langle i \rangle = 29^\circ$, which is practically the same as 
the median value that we obtain from the data in Table\,\ref{two_components},
of $i=27^\circ\pm 9^\circ$.

Similar disk inclinations have also been reported from the fit of disk models to double-peaked profiles in previous studies \citep{CeH89,EH94,Eracleous03}.

The fact that the majority of disks are observed with inclinations between 20$^\circ$ and 37$^\circ$ is supported
also by predictions of the simple model: if we consider that the maximum opening angle of the torus is $\theta_m\approx 37^\circ$,
then the fraction of Type 1 AGN observed with inclinations between 20$^\circ$ and 37\% should be $\sim$\,80\%,
while only $\sim$\,20\% should be observed with inclinations smaller than 20$^\circ$. This seems to be the
case of NGC\,5273 for which we obtain $i=17^\circ$, and of NGC\,3227, for which we fit the RMS
profile and obtain an inclination of  $i=16^\circ$. These low inclinations imply a larger value, $f\sim$12. Using this  $f$  value in equation \ref{eq:mass} would 
result in a mass for the SMBH 3 times larger than that obtained using the average $f$ value. Intriguingly (or coincidentally), the SMBH mass of NGC\,3227 obtained using reverberation mapping (RM) is a factor of ~2--3 lower than the dynamical mass estimates for this galaxy: \citet{Denney10} gives $M_{RM} = 6.0\times 10^6$M$_\odot$ (using $f=4.3$), while \citet{Hicks08} give $2.0\times 10^7$M$_\odot$  from gas kinematics, and \citet{Davies06} -- using stellar kinematics -- provide a 1$\sigma$ range spanning these two, i.e., $6.0\times 10^6-2.0\times 10^7$M$_\odot$.

\subsection{A relation between $f$ and the FWHM of the broad lines}

According to our model -- that applies for the case in which the BLR is dominated by emission from the disk --
 the adoption of an ensemble average value for $f$ is a good approximation for 80\% of Type 1 AGN
(even though there should be some variation between targets),
but for the 20\% cases of more pole-on AGNs the use of the constant $f$ implies an underestimation of the
SMBH mass.

Our model implies $f=1/\sin^2 i$ for the disk component. The facts that: (1) the parameters of the disk, and in particular, their inner, outer and
saturation radii do not vary much among different Type 1 AGN ($\sim$\,30\%, see Table\,\ref{two_components}); and (2) that the width (FWHM) of the line strongly depends on the inclination of the disk (see Fig.\,\ref{dp_profiles}) imply that there should be a relation between the FWHM of the broad profiles and $f$.


\begin{figure}
\centering
  \includegraphics[width=8.cm, scale=0.45]{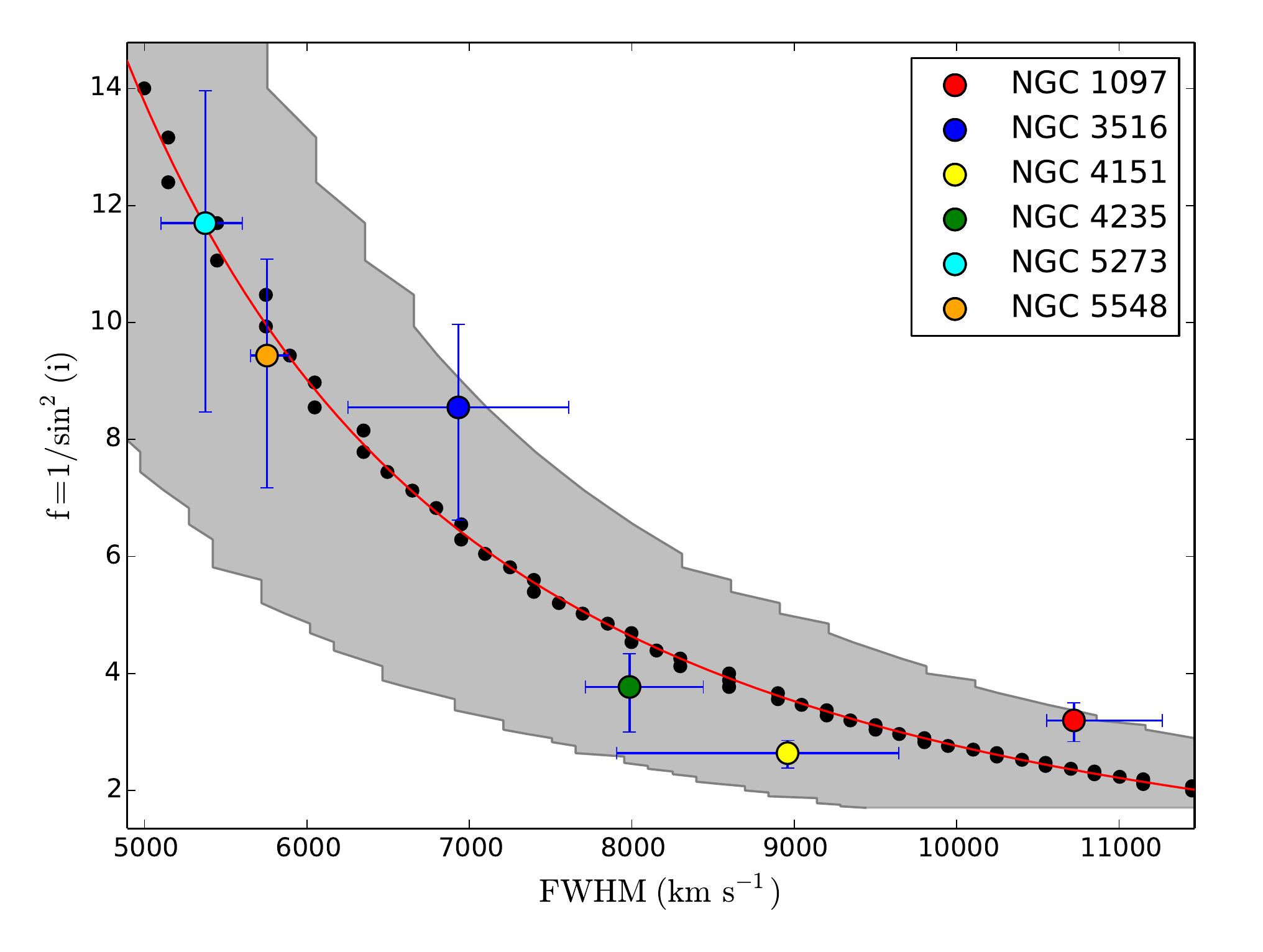}
  \caption{Relation between $f=1/\sin^2 i$ and the FWHM (km\,s$^{-1}$) of the broad line according to our average model (see last column of Table\,\ref{two_components}), 
represented by black
dots for inclinations in the range $10^\circ \le i \le 50^\circ$, and for the broad profiles of the Type 1 AGN of Table\,\ref{two_components} plus NGC\,1097 from Table\,\ref{n1097_table}. The grey band shows the range of the model values corresponding to the root-mean-square variations
of the parameter values listed in the last column of Table\,\ref{two_components}. The red line is a fit of the relation implied by the models.}
\label{f_vs_FW}
\end{figure}

We have used our model to obtain this relation between $f$ and FWHM of the profiles as follows. We have fixed the disk parameter values at the average values listed in the last column of Table\,\ref{two_components}, except for the inclination $i$. We have called this model ``average model". We have then varied the inclination $i$ between 10$^\circ$ and 50$^\circ$, to build a sequence of models as a function of $i$. For each model we measured the value of the FWHM (in km\,s$^{-1}$) and plotted $f$ as a function of FWHM in Fig.\ref{f_vs_FW} as black circles. We have then fitted the relation  between these two quantities with the function:

\begin{equation}
\label{eq:regression}
f = \frac{5.1\pm0.3\times10^9}{{\rm FWHM}^{2.30\pm0.04}}
\end{equation}

This relation is shown as a red line in Fig.\,\ref{f_vs_FW}. We have also included in Fig.\,\ref{f_vs_FW} the values obtained from the fit of the profiles of the 5 galaxies in Table\,\ref{two_components} plus that of NGC\,1097 in Table\,\ref{n1097_table}, shown as colored circles. The error bars correspond to the uncertainties in the inclinations listed in Table\,\ref{two_components} and in the FWHM values. The latter uncertainties were obtained as follows. While the listed FWHM values were obtained considering as  maximum flux the average between the peak fluxes of red and blue peaks or shoulders, the maximum FWHM  value was adopted as corresponding to a maximum flux equal to the flux of the lower peak and the minimum FWHM value as corresponding to a maximum flux equal do the flux of the higher peak. Fig.\,\ref{f_vs_FW}
shows that the measured FWHM values follow the relation given by equation \ref{eq:regression} within the uncertainties. 

Lower values of FWHM correspond to low disk inclinations, for which it is important to consider the thickness of the disk. For a ratio between a half thickness $H$ and disk radius $R$, $f$ becomes \citep{Collin06}:

\begin{equation}
\label{eq:H/R}
f = \frac{1}{{\rm(H/R)}^2+sin^2i} 
\end{equation}

Although this expression ensures that the value of $f$ will not diverge for low inclinations, we have checked that, for a typical value of $H/R=0.1$ \citep{Collin06}, the decrease in the value of $f$ by including the term $H/R$  is at most $\sim 10\%$ for the range of inclinations we are considering ($i\ge16^\circ$). 

The grey band in Fig.\,\ref{f_vs_FW} comprises the range of the model values corresponding to the root-mean-square variations of the parameter values listed in the last column of Table\,\ref{two_components}. The variations of the regression parameters of Equation~ \ref{eq:regression} allow the coverage of the grey band in Fig.\,\ref{f_vs_FW}.

Equation~ \ref{eq:regression} implies that, approximately,  $f\propto {\rm FWHM}^{-2}$.  We note that Equation~ \ref{eq:mass} is also consistent with this relation if the ratio $M_{\rm BH}/R_{\rm BLR}$ is constant. Our modeling, in which the profile shape depends on the inner, outer and maximum emissivity radia in units of gravitational radia imply exactly this: that the profile shape depends only on the ratio between the disk radia and the SMBH mass and that this ratio seems to be approximately constant. 

The fact that $f$ does not depend much on the ratio $M_{\rm BH}/R_{\rm BLR}$ is supported also by the following argument. From Equation \ref{eq:mass} we have: 

\begin{equation}
\label{eq:f}
 f \propto \frac{M_{BH}}{R_{\rm BLR}} \frac{1}{\Delta V^2}\mathrm{,}
\end{equation}

\noindent and we also know that $R_{\rm BLR}\propto L^{1/2}$ and that $L \propto \dot{M}$. The Eddington ratio is defined as $\dot{m}=\dot{M}/\dot{M}_{Edd}$ and $\dot{M}_{Edd} \propto M_{BH}$. So $\dot{M} \propto \dot{m}\dot{M}_{Edd} \propto \dot{m} M_{BH}$. Therefore

\begin{equation}
\label{eq:f_theory}
\frac{M_{BH}}{R_{\rm BLR}} \propto \frac{M_{BH}}{L^{1/2}} \propto \frac{M_{BH}}{(\dot{m}M_{BH})^{1/2}}
\end{equation}

\noindent and 

\begin{equation}
\label{eq:f_theory1}
f \propto \left(\frac{M_{BH}}{\dot{m}}\right)^{1/2} \frac{1}{\Delta V^2} 
\end{equation}

Thus, even for varying SMBH masses and accretion rates, the dependence of $f$ on these properties is weak, and, as a result,
$f$ depends mostly on the width of the profile: $f \propto \Delta V^{-2}$ or $f \propto {\rm FWHM}^{-2}$.




\subsection{Further implications for the Structure of AGN}
It has been argued that the broad double-peaked Balmer lines in LLAGN and radio galaxies originate
in the outer parts of an accretion
disk that is ionized by an ion torus \citep{CeH89} or other radiatively inefficient structure \citep{Narayan05,nemmen06}.
Such structure develops when the mass accretion rate to the SMBH drops below 
some minimum Eddington ratio, of order $\sim 1$\%, which is the case of LLAGN. 

LLAGN and Seyfert 1 nuclei have been compared to the two states 
observed in X-ray binaries \citep{ReM06, NeM08}: the low-luminosity hard state and the more luminous
soft state (or thermal state), respectively.
The thermal state is the one dominated by the geometrically thin, optically thick disk, analogous to more
lumious AGNs (Seyfert galaxies and quasars)
where the UV/optical emission is dominated by the ``big blue bump'' (BBB) from the accretion disk. The
low-luminosity hard state is characterized
by a harder X-ray spectrum and much lower thermal emission from the thin disk, and this is consistent
with the spectral energy distribution (SED) of LLAGN, which lacks the BBB component.

The fact that the double-peaked profiles are observed both in LLAGN and in Seyfert 1 galaxies suggest
that the inner structure of the low-luminosity
and high-luminosity AGN are not very different. The disk component seems to be always present. In LLAGN, 
the disk emission is the only component seen, while in more luminous Type 1
sources, such as Seyfert 1 nuclei and quasars, additional broad-line component that, 
since the velocity dispersion is smaller, probably originates in
more distant clouds than the gas that
produces the disk component. This is also supported by RM studies -- the fact the double-peaked characteristic pops out in the RMS spectrum, which shows only variable emission on RM time scales, suggests the lower-velocity gas ``filling-in" the peak is either optically thin to the continuum or else at a larger distance, where geometrical damping of the short time-scale RM variations occurs (much like the NLR).

The observation of the non-disk clouds leading to the 
additional broad component can be understood as due to the higher luminosity of the source;
at higher accretion rates there should be
more gas clouds in the vicinity of the nucleus, either inflowing or originating from an accretion disk wind that
gets enhanced with increased accretion rates.

%
%
%
%

\section{Conclusions}

We have used previously published results as well as fits of broad H$\alpha$ profiles in the spectra of nearby active Type 1 nuclei from the Palomar Survey of Nearby Galaxies to  argue that disk-like emission-line profiles -- double-peaked profiles originating in the outer parts of an accretion disk, at $\sim$\,1000 Schwarzdchild radii --  are ubiquitous in the BLR of these objects. 

We have shown that while ``pure" disk profiles are seen mostly  in LLAGN, in Seyfert 1 galaxies the profiles usually present an additional narrower component (FWHM\,$\sim$\,1000--3000\,km\,s$^{-1}$) well fitted with a Gaussian function. We attribute this component to non-disk BLR gas, usually moving at lower velocities. Our results support previous studies arguing that, while for LLAGN the accretion rate is low and most accreting gas is confined to the disk,  for the higher accretion rates that occur in Seyfert galaxies, one can also observe lower velocity ionized gas that is orbiting the SMBH farther from the disk, in inflow towards the disk or being ejected in a wind from the disk.

The disk component -- being closer to the ionizing source -- should be the most variable part of the profile, and indeed many reverberation mapping studies show that the $RMS$ spectrum reveals a double-peaked profile. 

While previous studies have already argued that the BLR is dominated by a flattened component, our study -- via the fit of a disk component to the BLR profiles  -- has allowed us to obtain the parameters of the disk, such as the inner and outer radii and the maximum emissivity radius in units of gravitational radii. We have found that these parameters do not vary by more than 30\% among the different sources, which leads to the conclusion that the main parameter that regulates the width of the double-peaked profile is the inclination of the disk relative to the plane of the sky.

In the cases for which the BLR emission is dominated by the disk component, the factor $f$ in the formula used to calculate the SMBH mass  $M_{\rm BH} = f \left(R_{\rm BLR} \Delta V^2/G\right)$ should be  $f=1/\sin^2 i$. The median inclination of the sources fitted with the disk model in our study, $i=27^\circ$, corresponds to $f=4.5$, very close to the most recent value of $\langle f \rangle = 4.3\pm 1.05$ \citep{Grier13B}, obtained in a completely independent way. But the range of inclinations we obtained, $17^\circ \le i \le 38^\circ$ implies that the $f$ factors range between $\sim$\,2.6 and $\sim$\,12. Thus, for the lowest  inclinations (narrower profiles), the SMBH masses obtained using a fixed value of $f$ are understimated by a factor of $\approx$3.

Having concluded that the main paramenter regulating the width of the BLR profiles -- in the disk-dominated cases -- is the inclination of the disk, our study has further allowed us to propose a relation (equation \ref{eq:regression}) between the value of $f$ and the FWHM of the broad profiles that could
lead to an improvement in the determination of SMBH masses in disk-dominated AGNs.

\acknowledgements
We thank the referee for a careful reading of the manuscript and valuable suggestions that helped to improve the paper. BMP is grateful for support by the NSF through grant AST-1008882 to The Ohio State University. KDD is supported by an NSF AAPF fellowship awarded under NSF grant AST-1302093.


\end{document}